# Solving hadron structures using the basis light-front quantization approach on quantum computers

Wenyang Qian,[1, 2, 3, *] Robert Basili,[2, 3, †] Soham Pal,[3, ‡] Glenn Luecke,[2, §] and James P. Vary[3, ¶]

[1]*Instituto Galego de Fisica de Altas Enerxias (IGFAE),
Universidade de Santiago de Compostela, E-15782 Galicia, Spain*
[2]*Department of Mathematics, Iowa State University, Ames, IA 50011, USA*
[3]*Department of Physics and Astronomy, Iowa State University, Ames, IA 50011, USA*

Quantum computing has demonstrated the potential to revolutionize our understanding of nuclear, atomic, and molecular structure by obtaining forefront solutions in non-relativistic quantum many-body theory. In this work, we show that quantum computing can be used to solve for the structure of hadrons, governed by strongly-interacting relativistic quantum field theory. Following our previous work on light unflavored mesons as a relativistic bound-state problem within the non-perturbative Hamiltonian formalism, we present the numerical calculations on simulated quantum devices using the basis light-front quantization (BLFQ) approach. We implement and compare the variational quantum eigensolver (VQE) and the subspace-search variational quantum eigensolver (SSVQE) to find the low-lying mass spectrum of the light meson system and its corresponding light-front wave functions as quantum states from ideal simulators, noisy simulators, and IBM quantum computers. Based on obtained quantum states, we evaluate the meson decay constants and parton distribution functions directly on the quantum circuits. Our calculations on the quantum computers and simulators are in reasonable agreement with accurate numerical solutions solved on classical computers when noises are moderately small, and our overall results are comparable with the available experimental data.

## I. INTRODUCTION

Quantum computing has emerged as a new method to simulate large-scale many-body quantum systems, which is a core challenge in the fields of chemistry and physics. Possessing the very same quantum mechanical nature that modern computational models of quantum systems seek to emulate (usually at great cost), quantum computing is considered a natural candidate for overcoming current resource-barriers faced by those models. In the current Noisy Intermediate-Scale Quantum (NISQ) era [1], our goal is to make full use of the available quantum computing resources to develop techniques for applications compatible with the noise of early quantum hardware. In addition to their necessity for achieving early quantum advantage, developing these techniques provide critical experience, insight, and points of comparison for later approaches to be applied with fault-tolerant quantum computation when it becomes available in the future.

Most current quantum computing applications for many-body systems rely on hybrid quantum-classical computers, such as the quantum approximate optimization algorithm (QAOA) [2] and the variational quantum eigensolver (VQE) [3–5]. The VQE algorithm was initially proposed to solve quantum chemistry problems [3, 6] and has now been applied to find the ground state energy of various nuclear systems [7–9]. In addition to obtaining ground states, the VQE algorithm can also be extended to solve for excited states [10–16]. In particular, the recently-proposed subspace-search variational quantum eigensolver (SSVQE) takes advantage of the orthogonality of the reference states and produces the specified spectroscopy in a single optimization step [15].

Light-front Hamiltonian approaches are particularly well-suited to quantum computing applications [8, 9, 17] as the eigenvalues of the Hamiltonian give rise to the complete spectroscopy and the light-front wave functions (LFWFs) enable direct access to physical observables. One such light-front Hamiltonian formalism is the basis light-front quantization (BLFQ) approach. The BLFQ approach utilizes basis functions to exploit the symmetry of the system to achieve a numerical advantage in high-performance computing [18], and it has already been successfully applied to many relativistic and strongly-interacting bound state systems [19–28]. In addition to obtaining the hadron mass spectroscopy, observables such as the decay constants, electromagnetic form factors, parton distribution functions, and parton distribution amplitudes can also be computed conveniently with LFWFs. In our previous work with BLFQ [28], we applied an effective light-front Hamiltonian to the light unflavored meson system to obtain the mass spectroscopy and physical observables for the low-lying states. By modifying the same effective Hamiltonian to a practical basis size allowed by quantum backends of currently-available quantum computers, we investigate the mass spectroscopy, decay constants, and parton distribution functions for the hadron system using the VQE and SSVQE approaches.

In this work, we will formulate the relativistic bound state problem of light meson systems and implement suit-able VQE and SSVQE programs to obtain the mass spectroscopy as well as other physical observables such as the decay constants. Our aim is to demonstrate a feasible path for solving the properties of hadrons on quantum computers which could lead to quantum advantage on future systems. At the same time, we will benchmark two of the available options, VQE and SSVQE, with simulations for currently-available systems.

* wqian@iastate.edu
† basiliro@iastate.edu
‡ soham@iastate.edu
§ grl@iastate.edu
¶ jvary@iastate.edu

We organize this paper as follows: In Sect. II, we introduce the effective Hamiltonian and BLFQ approach used to solve the light-front mass eigenvalue equation in the valence Fock sector of light-unflavored mesons. In Sect. III, we describe our implementation of the VQE and SSVQE methods to solve the bound-state eigenvalue problem, along with discussions of various encoding schemes and unitary ansatzes. In Sect. IV, we present the results of the mass spectroscopy, decay constants, and parton distribution functions for the selected states using VQE and SSVQE approaches on the light meson systems via quantum simulations and compare them to experiments and with the exact results obtained using classical methods. In Sect. V, we summarize our results and discuss possible future developments.

## II. EFFECTIVE HAMILTONIAN AND BASIS FUNCTION REPRESENTATION

### A. The Hamiltonian

We adopt the formalism and effective Hamiltonian for light unflavored meson systems proposed in a previous work [28]. The effective light-front Hamiltonian in a convenient but mixed representation (both momenta and coordinates are employed) reads

$$
\begin{aligned}
H_{\text{eff}} &\equiv P^+ P^-_{\text{eff}} - \boldsymbol{P}^2_\perp \\
&= \frac{\boldsymbol{k}^2_\perp + m_q^2}{x} + \frac{\boldsymbol{k}^2_\perp + m_{\bar{q}}^2}{1-x} + \kappa^4 x(1-x) \boldsymbol{r}^2_\perp \\
&- \frac{\kappa^4}{(m_q + m_{\bar{q}})^2} \partial_x (x(1-x)\partial_x) \\
&- \frac{C_{\text{F}} 4\pi \alpha_s(Q^2)}{Q^2} \bar{u}_{s'}(k') \gamma_\mu u_s(k) \bar{v}_{\bar{s}}(-k) \gamma^\mu v_{\bar{s}'}(-k'), \quad (1)
\end{aligned}
$$

where $m_q$ ($m_{\bar{q}}$) is the mass of the quark (anti-quark), $\kappa$ is the strength of the confinement, $\boldsymbol{k}_\perp$ ($-\boldsymbol{k}_\perp$) is the relative momentum of the quark (anti-quark), $x$ ($1-x$) is the longitudinal momentum fraction of the quark (anti-quark), and $\boldsymbol{r}_\perp$ is the transverse separation of the quark and the anti-quark. The first two terms are the light-front kinetic energy of the quark and the anti-quark. The third term adopts the light-front anti-de Sitter/quantum chromodynamics (AdS/QCD) soft-wall potential [29, 30] to implement the transverse confinement. The fourth term serves as the longitudinal confinement [20] by supplementing the transverse confinement to form a 3-dimensional spherical confinement potential in the non-relativistic limit. The fifth and last term is the one-gluon exchange based on one-loop perturbative QCD (pQCD) [21] to produce spin-dependent interactions at short distance, where $C_{\text{F}} = (N_c^2 - 1)/(2N_c) = 4/3$ is the color factor with $N_c = 3$, and $Q^2$ is the average 4-momentum square carried by the exchanged gluon. It is important to note that the contribution of the pseudoscalar interaction in the original paper is neglected, since we will be using very limited basis spaces to perform quantum simulation in this work.

With the Hamiltonian defined in Eq. (1), the mass spectrum and wave functions can be obtained directly by solving the light-front eigenvalue equation

$$H_{\text{eff}} |\psi(P, j, m_j)\rangle = M^2 |\psi(P, j, m_j)\rangle, \quad (2)$$

where $P = (P^-, P^+, \boldsymbol{P}_\perp)$ is the four momentum of the hadron in light-front coordinates (Appendix A), $j$ is the total angular momentum, $m_j$ is the magnetic projection, and $M$ is the mass of the hadron. Working within the leading $|q\bar{q}\rangle$ Fock sector, the meson state is written as

$$
\begin{aligned}
|\psi(P, j, m_j)\rangle &= \sum_{s,\bar{s}} \int \frac{\mathrm{d}x}{2x(1-x)} \int \frac{\mathrm{d}^2 \boldsymbol{k}_\perp}{(2\pi)^3} \psi^{m_j}_{s\bar{s}}(\boldsymbol{k}_\perp, x) \\
&\times \frac{1}{\sqrt{N_c}} \sum_{i=1}^{N_c} b^\dagger_{si}(xP^+, \boldsymbol{k}_\perp + x\boldsymbol{P}_\perp) \\
&\times d^\dagger_{\bar{s}i}((1-x)P^+, -\boldsymbol{k}_\perp + (1-x)\boldsymbol{P}_\perp) |0\rangle, \quad (3)
\end{aligned}
$$

where $\psi^{m_j}_{s\bar{s}}(\boldsymbol{k}_\perp, x)$ is the light-front wave function (LFWF) of the hadron, $s$ and $\bar{s}$ represent the spin of the quark and anti-quark, and the quark and anti-quark creation operators $b^\dagger$ and $d^\dagger$ satisfy the canonical anti-commutation relations,

$$
\begin{aligned}
&\{b_{si}(p^+, \boldsymbol{p}_\perp), b^\dagger_{s'i'}(p'^+, \boldsymbol{p}'_\perp)\} \\
&= \{d_{si}(p^+, \boldsymbol{p}_\perp), d^\dagger_{s'i'}(p'^+, \boldsymbol{p}'_\perp)\} \\
&= 2p^+ (2\pi)^3 \delta(p^+ - p'^+) \delta^2(\boldsymbol{p}_\perp - \boldsymbol{p}'_\perp) \delta_{ss'} \delta_{ii'}.
\end{aligned} \quad (4)
$$

### B. Basis Function Representation

To solve the eigenvalue equation in Eq. (2), we use basis light-front quantization (BLFQ) approach, where the Hamiltonian is diagonalized within a chosen basis function representation [18]. In this work, we use the same basis function adopted in Ref. [28] which are convenient basis functions for the relative motion dynamics. That is, the center-of-mass motion does not appear since $H_{\text{eff}}$ acts only on the relative motion of the quark and anti-quark. Explicitly, we expand the LFWF $\psi^{m_j}_{s\bar{s}}(\boldsymbol{k}_\perp, x)$ into the transverse and longitudinal basis functions with coefficients $\tilde{\psi}^{m_j}_{s\bar{s}}(n, m, l)$:

$$\psi^{m_j}_{s\bar{s}}(\boldsymbol{k}_\perp, x) = \sum_{nml} \tilde{\psi}^{m_j}_{s\bar{s}}(n, m, l) \phi_{nm}(\frac{\boldsymbol{k}_\perp}{\sqrt{x(1-x)}}) \chi_l(x), \quad (5)$$

where

$$\phi_{nm}(\boldsymbol{q}_\perp) = \frac{1}{\kappa}\sqrt{\frac{4\pi n!}{(n+|m|)!}}\left(\frac{q_\perp}{\kappa}\right)^{|m|}$$
$$\times e^{-\frac{q_\perp^2}{2\kappa^2}} L_n^{|m|}\left(\frac{q_\perp^2}{\kappa^2}\right) e^{im\theta_q}, \quad (6)$$

$$\chi_l(x;\alpha,\beta) = x^{\frac{\beta}{2}}(1-x)^{\frac{\alpha}{2}} P_l^{(\alpha,\beta)}(2x-1)$$
$$\times \sqrt{4\pi(2l+\alpha+\beta+1)}$$
$$\times \sqrt{\frac{\Gamma(l+1)\Gamma(l+\alpha+\beta+1)}{\Gamma(l+\alpha+1)(\Gamma(l+\beta+1)}}. \quad (7)$$

In the transverse direction, we use the 2-dimensional harmonic oscillator function $\phi_{nm}(\boldsymbol{q}_\perp)$, where $\boldsymbol{q}_\perp \triangleq \boldsymbol{k}_\perp/\sqrt{x(1-x)}$, $q_\perp = |\boldsymbol{q}_\perp|$, $\theta_q = \arg \boldsymbol{q}_\perp$, and $L_n^a(z)$ is the generalized Laguerre polynomial. The confining strength $\kappa$ serves as the harmonic oscillator scale parameter. Integers $n$ and $m$ represent the principal quantum number for radial excitations and the orbital angular momentum projection quantum number, respectively. In the longitudinal direction, we use the basis function $\chi_l(x;\alpha,\beta)$, where $l$ is the longitudinal quantum number, $P_l^{(\alpha,\beta)}(2x-1)$ is the Jacobi polynomial, $\alpha = 2m_{\bar{q}}(m_q + m_{\bar{q}})/\kappa^2$ and $\beta = 2m_q(m_q + m_{\bar{q}})/\kappa^2$. In particular, the basis function is constructed to preserve the magnetic projection of total angular momentum, $m_j = m + s + \bar{s}$.

The basis function approach offers a numerically efficient way to discretize the Hamiltonian. In practice, the transverse and longitudinal basis functions are truncated to their respective transverse cutoff $N_{\max}$ and longitudinal cutoff $L_{\max}$:

$$2n + |m| + 1 \leq N_{\max}, \quad 0 \leq l \leq L_{\max}. \quad (8)$$

$N_{\max}$ controls the total allowed oscillator quanta in the system and $L_{\max}$ controls the longitudinal basis resolution. The BLFQ Hamiltonians have been demonstrated to produce results for mass spectroscopy and other observables that scale well with the energy cutoffs $N_{\max}$ and $L_{\max}$ [21]. The exact spectra and LFWFs correspond to results without cutoffs, i.e. the infinite matrix limit or the continuum limit. It is anticipated that quantum computers will someday surpass classical computers and more closely approach the continuum limit.

## III. VARIATIONAL QUANTUM EIGENSOLVER

Having defined the eigen-problem and its basis representation, we are ready to describe the variational quantum eigensolver approaches that we adopt to perform quantum simulations.

### A. Variational Quantum Eigensolver

Given a Hermitian matrix $H$ with an unknown minimum eigenvalue $\lambda_{\min}$ associated with the eigenstate $|\psi_{\min}\rangle$, the variational principle provides an estimate $\lambda_\theta$ upperbounding $\lambda_{\min}$,

$$\lambda_{\min} \leq \lambda_{\vec{\theta}} \equiv \langle \psi(\vec{\theta})| \hat{H} |\psi(\vec{\theta})\rangle, \quad (9)$$

where $\vec{\theta}$ is a list of parameters, and $|\psi(\vec{\theta})\rangle$ is a parameterized eigenstate associated with $\lambda_{\vec{\theta}}$.

The variational quantum eigensolver (VQE) [3, 31] is a hybrid computational approach consisting of a quantum part and a classical part. In the quantum part, a prepared parameterized quantum circuit, represented by the unitary $\hat{U}(\vec{\theta})$, is applied to an initial state, $|\psi_0\rangle$, to obtain a final state, $|\psi(\vec{\theta})\rangle \equiv \hat{U}(\vec{\theta})|\psi_0\rangle$, that estimates $|\psi_{\min}\rangle$. In the classical part, the estimate is iteratively optimized using a classical optimizer by changing the parameter $\vec{\theta}$ in each iteration to minimize the expectation value of the Hamiltonian, $\langle \psi(\vec{\theta})| \hat{H} |\psi(\vec{\theta})\rangle$. The algorithm terminates when a specified numerical tolerance or a maximum allowed iteration is achieved.

Specifically, the general procedure for solving an eigenvalue Hamiltonian problem with the VQE approach can be divided as follows:

1. Select the Hamiltonian $\hat{H}$ for the targeted physical system and a suitable mapping scheme onto a set of qubits.

2. Pick an initial state $|\psi_0\rangle$ and a parameterized unitary ansatz $\hat{U}(\vec{\theta})$ for state evolution.

3. Apply the unitary ansatz to the initial state to obtain the final state $|\psi(\vec{\theta})\rangle = \hat{U}(\vec{\theta})|\psi_0\rangle$ and measure the cost function, or the expectation value of the Hamiltonian $\langle \psi(\vec{\theta})| \hat{H} |\psi(\vec{\theta})\rangle$. (quantum computer)

4. Optimize the parameter $\vec{\theta}$ by minimizing the cost function which is the expectation value. (classical computer)

Step 1 and 2 can usually be prepared before the actual VQE iterations. Each measurement in step 3 is ideally performed on the quantum computer by running repeated instances (or 'shots') of the quantum circuit to sample the probability distribution of the final state, and step 4 is computed on the classical computer using various optimizers available. In the end, both step 3 and 4 are repeated over many iterations to obtain the final optimal parameter $\vec{\theta}^*$. Optimization can be made in each step to improve the overall performance.

### B. Subspace-search Variational Quantum Eigensolver

The VQE approach can be further extended to the subspace-search variational quantum eigensolver

(SSVQE) [15] to find excited states of the system by restricting the subspace of unitary evolution and by considering a different set of cost functions. One variant of this approach is the weighted SSVQE. Instead of minimizing a single expectation value of the Hamiltonian, weighted SSVQE considers the cost function to be a weighted sum of a set of expectation values of the Hamiltonian, each measured from an orthogonal initial reference state after the unitary evolution. To find up to the $k$th excited states, the algorithm is as follows:

1. Select the Hamiltonian $\hat{H}$ for the targeted physical system and a suitable mapping scheme onto a set of qubits.

2. Pick a set of mutually orthogonal initial states $\{|\psi_i\rangle\}_{i=0}^{k}$, and a parameterized unitary ansatz $\hat{U}(\vec{\theta})$ acting on these states.

3. Apply the unitary ansatz to each state and measure their expectation values, $\vec{E} = (E_0, E_1, \cdots, E_k)$, where $E_i = \langle \psi_i(\vec{\theta}) | \hat{H} | \psi_i(\vec{\theta}) \rangle$. (quantum computer)

4. Optimize the parameter $\vec{\theta}$ by minimizing the cost function $\mathcal{C}_{\vec{\omega}}(\vec{\theta}) = \vec{\omega} \cdot \vec{E}$, where $\vec{\omega}$ is a straightly decreasing weight vector prioritizing lower-lying states ($\omega_i > \omega_j$ for $i < j$). (classical computer)

In particular, if we just want to look for the $k$th excited state, we can also modify the weight vector such that $0 < w_k < 1$ and $w_i = 1$ for all $0 \leq i < k$. With a single optimization procedure, the weighted SSVQE is capable of obtaining the specified low-lying spectrum exactly. However, extra quantum computing resources are needed to evaluate all $k$ expectation values within each iteration step. In the following subsections, we further describe each step for the VQE and SSVQE algorithms.

### C. Mapping the Hamiltonian to Qubits

Here, we will discuss two suitable encoding schemes to map a hadronic Hamiltonian to qubits. Using second quantization, a generic Hamiltonian $\hat{H}$ is represented in terms of the creation operators ($\hat{a}^\dagger$) and annihilation operators ($\hat{a}$):

$$\hat{H} = \hat{H}_1 + \hat{H}_2 + \cdots$$
$$= \sum_{ij} h_{ij} \hat{a}_i^\dagger \hat{a}_j + \frac{1}{4} \sum_{ijkl} h_{ijkl} \hat{a}_i^\dagger \hat{a}_j^\dagger \hat{a}_k \hat{a}_l + \cdots, \quad (10)$$

where $\hat{H}_1$ represents the single excitation interactions, $\hat{H}_2$ represents the double excitation interactions and so forth. In single-particle fermion states, $h_{ijkl}$ has a sign change under the interchange of either the first two or the last two indexes. In this work, we are working with relative coordinate representation of the meson system where the quantum properties of identical particles does not play a role. Instead, the creation operators can be viewed as symbolizing the creation of a specified mode of relative motion. For our meson system in the quark/antiquark space, we restrict ourselves to the first term in Eq. (10) and its coefficient $h_{ij}$ corresponds to the matrix elements in the basis representation of the BLFQ Hamiltonian. All modes accessible in the system are created by a corresponding BLFQ creation operator acting on the vacuum.

To solve the Hamiltonian problem on quantum computers in practice, we need to encode the physical states as well as any unitary operators onto the qubits. Various mapping schemes are proposed, such as the Jordan-Wigner (JW) representation [32], Bravyi-Kitaev representation [33], a compact representation [9, 17], and so forth. Here, we focus on the so-called direct encoding as described by the JW representation [32] and the compact encoding according to the Hilbert-Schmidt decomposition [34].

In the JW representation, we map directly from the fermionic operators to the many-state Pauli spin matrices. Specifically, we write the creation and annihilation operators as

$$\hat{a}_j^\dagger = \bigotimes_{i=1}^{j-1} Z_i \otimes \frac{X_j - iY_j}{2}, \quad (11)$$

$$\hat{a}_j = \bigotimes_{i=1}^{j-1} Z_i \otimes \frac{X_j + iY_j}{2}, \quad (12)$$

where $X_i, Y_i, Z_i$ are the Pauli-X, Y, Z matrices acting on the corresponding $i$-th qubit (Appendix B). With this construction on many-qubit states, the canonical commutation relations for fermions, $\{\hat{a}_i, \hat{a}_j\} = 0$ and $\{\hat{a}_i, \hat{a}_j^\dagger\} = \delta_{ij}$, are satisfied [35]. The substitution of Eq. (11) and Eq. (12) into Eq. (10) gives rise to the desired qubitized Hamiltonian operator $H_q$ acting on the many-qubit state. In the JW encoding, we need $N$ qubits to encode an N-by-N Hamiltonian matrix properly, where $N$ is always a power of two, $N = 2^n$.

In the compact representation, using the orthogonal basis formed by the Pauli strings under trace, we can decompose an N-by-N (or $2^n$-by-$2^n$) Hamiltonian matrix $H$ into its qubitized form by

$$H_q = \frac{1}{N} \sum_{\alpha=1}^{N^2} \text{Tr}(P_\alpha H) \cdot P_\alpha, \quad (13)$$

where $P_\alpha = \otimes_{k=1}^{n} \sigma_k$ is an $n$-qubit Pauli string and $\sigma_k \in \{I_k, X_k, Y_k, Z_k\}$ is a Pauli matrix acting on the $k$-th qubit. Since $\text{Tr}(P_\alpha P_\beta) = 2^n \delta_{j,k} = N\delta_{j,k}$ for any two $P_\alpha, P_\beta$, and there exists $4^n = N^2$ distinct $P_\alpha$, the set of all the Pauli strings form an orthogonal basis under trace for any N-by-N matrix. In this encoding, we only need $n = log_2(N)$ qubits to encode an N-by-N matrix properly.

With either encoding scheme, the original Hamiltonian is now expressed as a sum of Pauli strings acting on the many-qubit state,

$$\hat{H} = \sum_{ij} h_{ij} \hat{a}_i^\dagger \hat{a}_j \to H_q = \sum_\alpha c_\alpha P_\alpha, \qquad (14)$$

where $P_i$ is a Pauli string whose length depends on the encoding and $c_\alpha$ is its respective coefficient. It is worth pointing out that the number of Pauli string terms could pessimistically scale as $4^n$ with the number of qubits. In practice, however, measurement reduction techniques may be adopted to significantly reduce the number of expectation evaluations by grouping the Pauli terms into commuting collections for simultaneous measurement [36–40]. In fact, finding the necessary number of measurements is equivalent to the NP-hard minimum clique cover problem, where heuristic approximate solutions can be used [37]. Additionally, as nuclear physics Hamiltonians are often sparse matrices, efficient Hamiltonian encoding strategies [41, 42] can also be used to directly reduce the number of Pauli terms. Lastly, adaptive and intelligent optimizers can further decrease the total number of measurements in the optimization loop [43, 44]. In the simulation results of this work, we always group commuting Pauli terms to minimize the cost of quantum measurements.

### D. Unitary Ansatzes

We now need to select a suitable ansatz $\hat{U}(\vec{\theta})$ to evolve the initial state to some final states on the quantum circuit, where $\vec{\theta}$ contains all the parameterizations of the ansatz. For JW encoding, the Unitary Coupled Cluster (UCC) [45, 46] ansatz, based on traditional coupled cluster methods, has emerges as one of the popular ansatzes. In general, the variational UCC ansatz is defined as

$$\hat{U}(\vec{\theta}) = e^{\hat{T}(\vec{\theta}) - \hat{T}^\dagger(\vec{\theta})}, \qquad (15)$$

$$\hat{T}(\vec{\theta}) = \sum_{i=1}^n \hat{T}_i(\vec{\theta}) = \hat{T}_1(\vec{\theta}) + \hat{T}_2(\vec{\theta}) + \cdots, \qquad (16)$$

where the excitation operator $\hat{T}$ can be written as a sum of single excitation $\hat{T}_1$, double excitations $\hat{T}_2$, and higher order excitations, each corresponding to its respective term in the second quantized form of the Hamiltonian. Specifically for our work within single excitation,

$$\hat{T}_1(\vec{\theta}) = \sum_{\substack{r \in \text{occ} \\ p \in \text{virt}}} \theta_p^r \hat{a}_p^\dagger \hat{a}_r, \qquad (17)$$

where the *occ* and *virt* subspaces are defined as the occupied and unoccupied qubit orbital (or mode in our specific application) in the reference state, and $\theta_p^r$ are the expansion coefficients. The variational UCC ansatz allows one to span the allowed Hilbert space entirely starting from the given initial state. According to JW representation, we can show that each pair of Hermitian operators in $\hat{T}_1$ (for $i > j$) as

$$\hat{a}_i^\dagger \hat{a}_j - \hat{a}_j^\dagger \hat{a}_i = \frac{i}{2} \bigotimes_{a=j+1}^{i-1} Z_a (Y_j X_i - X_j Y_i) \qquad (18)$$

and the variational ansatz can be conveniently represented as a sum of Pauli strings $P_\alpha$ with real coefficients $c_\alpha$,

$$\hat{U}(\vec{\theta}) = e^{i \sum_\alpha c_\alpha P_\alpha}. \qquad (19)$$

The UCC unitary ansatz can be approximated via trotterization [46, 47],

$$\hat{U}(\vec{\theta}) \approx \hat{U}_{\text{Trot}}(\vec{t}) = \left( \Pi_\alpha e^{i \frac{c_\alpha}{\rho} P_\alpha} \right)^\rho, \qquad (20)$$

where $\rho$ is the trotter number. In practice, the trotter number is usually quite small. We use $\rho = 1$ in this work. A partial quantum circuit of the $\hat{U}(\vec{\theta})$ is shown in Fig. 1. The UCC ansatz takes only a couple of parameters within single excitations but may result in a rather large circuit depth and agnostic to device connectivity.

Recently, another promising type of ansatz to consider is the so-called hardware efficient ansatz (HEA) [31], where its circuit is composed of alternating single-qubit rotation layers and entanglement layers. The parameters of the HEA are exactly the Euler angles specified in each rotation layer. The entanglement layer can have various many-qubit gate implementations to generate sufficient entanglement. The HEA is a heuristic ansatz and allows us to design quantum circuits that best match a given quantum hardware layout. For the same reason, it may also be difficult to achieve the same accuracy using the HEA as one achieves with the previous problem-inspired UCC ansatz.

Another complication with the HEA is the potential risk of vanishing gradient or barren plateaus [49], for some type of HEAs especially when their initial parameters are randomly chosen. Many solutions have been proposed to resolve this issue such as using selected initial points [50], revising cost function [51], and designing trainable and expressible ansatzes [52]. While preparing this work, we considered implementing the alternating layered ansatz (ALT) and tensor product ansatz (TEN) approaches mentioned in Ref. [52] to prevent a vanishing gradient, but found their performances are similar to that of HEAs at the relatively small problem scales being considered here. We anticipate that the differences of these approaches from HEAs will become more apparent at larger problem scales, and thus leave their further consideration for a future work where such scales are considered.

Therefore, in this work, we will focus on the hardware efficient SU(2) two-local ansatz, which is provided by the native `EfficientSU2` class [53] from Qiskit. For a two-qubit HEA with a single repetition layer, the circuit is shown in Fig. 2. Since the number of parameters in the HEA scales linearly with both the number of qubits and the number of repeated layers, the HEA usually takes significantly more parameters to be optimized. To some extent, this can be regarded as a trade-off between the number of parameters and the depth of the quantum circuit, which can be particularly advantageous for quantum hardware that is currently-available.

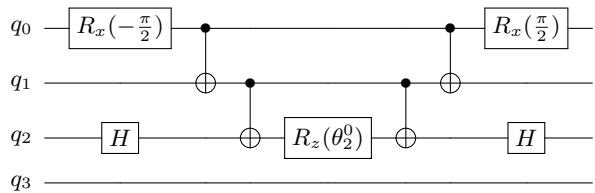

FIG. 1. Quantum circuit of one of the possible terms, $e^{\theta_2^0 \hat{a}_2^\dagger \hat{a}_0} \to e^{i\theta_2^0 X_2 Z_1 Y_0}$, in the four-qubit UCC ansatz [48] within first excitation. Here, the occupied qubit is $q_0$ and the virtual qubit is $q_2$. The full four-qubit UCC circuit starting from a single occupied state, for example $q_0$, consists of six such subcircuits in total, which makes it a very deep circuit. Note the extra factor of two, in Eq. (18), is absorbed into the parameter.

It should be noted that noise-induced barren plateaus [54] could still be present in a generic ansatz, for both the UCC and HEA ansatzes considered in our case, as the gradient vanishes exponentially in the number of qubits when the ansatz depth grows linearly. However, in our noisy simulations, this phenomenon is barely observed and we will defer further investigation to a future work.

### E. Measurement and Optimization

With a parameterized unitary ansatz that takes an initial state of our choice to a final state, we are able to measure the expectation value of the Hamiltonian. Since the Hamiltonian consists of many different sub-terms, we collect commuting sets of sub-terms to measure them separately on the quantum computer. Each measurement often takes thousands of shots in order to obtain a histogram of the final quantum state. Post-measurement operations are appended as needed, such as applying a Hadamard gate or Rotation-Y gate to change basis when measuring Pauli-Y and Pauli-X spin matrices, respectively. In the end, we obtain a single numerical value that is the best approximate of the expected eigenvalue by summing up all relevant expectations of the sub-terms in the Hamiltonian.

The measured eigenvalue is passed onto the classical computer and we use various optimizers to update the parameters for the next iteration. In this work, we used the Constrained Optimization BY Linear Approximation (COBYLA) [55–57] optimizer, the Limited-memory Broyden-Fletcher-Goldfarb-Shanno Bound (LBFGSB) [58–60] optimizer, and Sequential Least SQuares Programming (SLSQP) [61] optimizer from `scipy.optimize` library. We also used the noise-resilient Simultaneous Perturbation Stochastic Approximation (SPSA) [62, 63] optimizer and the Quantum Natural SPSA (QNSPSA) [64] optimizer from `qiskit.algorithms.optimizers` library. In general, we find the LBFGSB optimizer most suitable for exact simulations and the SPSA optimizer best-performing for noise-free, noisy and real quantum simulations [65, 66] in obtaining the lowest cost expectation. The gradient-free COBYLA optimizer can be very useful across all simulations primarily due to its short iterations for convergence and resilience to low noises. Shot-frugal optimizers such as Rosalin (Random Operator Sampling for Adaptive Learning with Individual Number of shots) [44] that performs weighted random sampling of the cost Hamiltonian could potentially improve the simulation result and runtime when our Hamiltonian system scales in the future.

After iterated optimizations, one is expected to get the converged parameters, the expectation values, and, most importantly, the final state resulting from the given unitary ansatz. It is crucial to run the quantum simulation multiple times as the initial starting parameters can have a large impact on the optimization outcome. Different optimizers are also sensitive to different initial parameters. In each of our results below, we have performed multiple simulations and only presented the simulation result with the lowest value of the final cost function.

## IV. NUMERICAL RESULTS

### A. Qubitized Hamiltonian

In this work, we use the BLFQ Hamiltonian obtained from [28] in a relatively smaller basis that is more suitable to currently-available quantum computing resources. We work within the SU(2) isospin symmetric limit such that the anti-quark and the quark masses are identical. The values for the number of quark flavors $N_f$ and the strong coupling coefficient are directly taken from the previous work. We consider the three smallest but physically significant choices of the basis sizes: $(N_{\max}, L_{\max}) = (1, 1)$, $(N_{\max}, L_{\max}) = (4, 1)$, and $(N_{\max}, L_{\max}) = (4, 3)$. Respectively, they correspond to matrix dimensions of 4, 16, and 32. Variation in $N_{\max}$ represents the sensitivity in radial excitations while variation in $L_{\max}$ probes the longitudinal excitations. The quark mass $m_q$ and the confining strength $\kappa$ are obtained by fitting the experimental mass of $\rho(770)$ meson from the Particle Data Group (PDG) [67] at their respective cutoffs, where the observed difference in $m_q$ can be viewed as the correction of their effective masses. The model parameters are summarized in Table I and they differ slightly from values in Ref. [28].

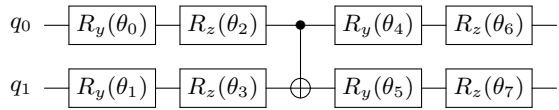

FIG. 2. Quantum circuit for the two-qubit HEA with a single repetition layer using `EfficientSU2`.

According to the basis representation of Eq. (6) and Eq. (7), the BLFQ Hamiltonian matrix $H_{\text{eff}}^{(1,1)}$ for $N_{\max} = 1$ and $L_{\max} = 1$ is obtained as follows

$$\begin{pmatrix} 568487 & 0 & 25428 & 0 \\ 0 & 1700976 & 0 & -15767 \\ 25428 & 0 & 568487 & 0 \\ 0 & -15767 & 0 & 1700976 \end{pmatrix}, \quad (21)$$

where each matrix element is rounded to the nearest integer and expressed in units of MeV$^2$. Directly solving the Hamiltonian by matrix diagonalization on classical computers produces four eigenvalues, whose square roots correspond to the four states in the mass spectrum:

$$737 \text{ MeV}, \quad 771 \text{ MeV}, \quad 1298 \text{ MeV}, \quad 1310 \text{ MeV},$$

where the second mass is fitted exactly with the experimental $\rho(770)$ meson mass.

To map the basis states of the Hamiltonian to qubits, we identify the available basis states for $N_{\max} = 1$ and $L_{\max} = 1$ as follows in Table. II. Together with Eq. (11), Eq. (12) and Eq. (13), we obtain the directly-encoded Hamiltonian operator $H_{\text{direct}}^{(1,1)}$ on 4 qubits and the compactly-encoded Hamiltonian operator $H_{\text{compact}}^{(1,1)}$ on 2 qubits respectively for the same Hamiltonian matrix $H_{\text{eff}}^{(1,1)}$,

$$\begin{aligned} H_{\text{direct}}^{(1,1)} &= 2269462 \, \text{IIII} - 284243 \, (\text{ZIII} + \text{IIZI}) \\ &\quad - 850488 \, (\text{IZII} + \text{IIIZ}) \\ &\quad + 12714 \, (\text{XZXI} + \text{YZYI}) \\ &\quad - 7883 \, (\text{IXZX} + \text{IYZY}), \end{aligned} \quad (22)$$

$$\begin{aligned} H_{\text{compact}}^{(1,1)} &= 1134731 \, \text{II} - 566245 \, \text{IZ} \\ &\quad + 4831 \, \text{XI} + 20598 \, \text{XZ}, \end{aligned} \quad (23)$$

where each qubit operator is written as a sum of Pauli strings with the leading Pauli matrix acting on the qubit with the highest index and so on. For the larger Hamiltonians of $N_{\max} = 4$ and $L_{\max} = 1$ or 3, we will only focus on their compactly-encoded operators $H_{\text{compact}}^{(4,1)}$ and $H_{\text{compact}}^{(4,3)}$ due to the intense computational resources needed for direct encoding. In particular, we include $H_{\text{compact}}^{(4,1)}$ in Appendix E along with its basis encoding in Table. VII for comparison.

### B. Spectroscopy

#### 1. Results of VQE

With the Hamiltonian mapped onto the qubits, we first show the results of using the VQE algorithm to compute the ground state energy for the Hamiltonian at $N_{\max} = L_{\max} = 1$ in Fig. 3. The left panel of the figure shows the results using the directly-encoded 4-qubit operator $H_{\text{direct}}^{(1,1)}$ from Eq. (22), and the right panel shows the results using compactly-encoded 2-qubit operator $H_{\text{compact}}^{(1,1)}$ from Eq. (23). For the direct encoding, we use the 4-qubit UCC ansatz with one single trotterization and set $|0001\rangle$ as the initial state (or set $q_0$ as the occupied qubit), where part of the circuit is shown in Fig. 3. For the compact encoding, we use the 2-qubit HEA with one single repetition layer as shown in Fig. 3 and set $|00\rangle$ as the initial state. The detailed summary of each quantum circuit is presented in Table. III. Within each set of VQE applications, we used both the statevector (SV) simulator from `aer.StatevectorSimulator` and the QASM simulator from `aer.QasmSimulator` to simulate the quantum apparatus and calculate the ground state energy. The SV simulator is an ideal quantum circuit statevector simulator that returns the quantum state exactly, which is useful for debugging and theoretical testing; the QASM simulator (the main Qiskit Aer backend) emulates the execution of the quantum circuit on an ideal quantum device and returns measurement counts with statistical uncertainty from designated number of shots at the end of the simulation. From Fig. 3, both the quantum simulator results (SV and QASM) are in good agreement with the exact ground state energy obtained by diagonalizing the original Hamiltonian matrix from Eq. (21).

In addition to classical simulators, we include quantum computer results using IBM's 5-qubit superconducting processor, IBMQ Manila, in Fig. 3 to obtain the ground state energy with the help of the newly-proposed Qiskit runtime library, `VQEClient`. For both quantum simulations, we use QNSPSA optimizers for the expectation value obtained from 8,192 shots at each step of the optimization. To mitigate the readout error, we perform the complete re-calibration every 30 minutes and apply the measurement correction filter to all of our measurements using `CompleteMeasFitter` from the Qiskit mitigation library. For the compact encoding, we use a 2-qubit hardware efficient quantum circuit with a depth of 11 and we are able to obtain the ground state energy in agreement with the exact mass eigenvalue. On the other hand, with direct encoding and the UCC ansatz, we did not obtain a converged result as expected, because the full UCC circuit at a depth of 67 overwhelms the maximal coherence length allowed by the IBM Manila backend with a quantum volume of 32. To resolve this problem on quantum computers, there are many solutions, such as picking a quantum backend with longer coherence time or preparing a specialized ansatz [9, 68] made for the Hamiltonian, but they are not within the scope of this work. For today's NISQ devices, it seems more advantageous to shift the computational burden, i.e. the number of parameters, on the classical optimizers, than to have a lengthy quantum circuit.

TABLE I. Model parameters of the BLFQ Hamiltonian. All three Hamiltonians $H_{\text{eff}}^{(1,1)}$, $H_{\text{eff}}^{(4,1)}$, and $H_{\text{eff}}^{(4,3)}$ use the quark mass $m_q$ and the confining strength $\kappa$ fit to spectra as described in the text at the specified basis truncations.

|  | $N_{\text{f}}$ | $\alpha_{\text{s}}(0)$ | $\kappa$ (MeV) | $m_q$ (MeV) | $N_{\max}$ | $L_{\max}$ | Matrix dimension |
|---|---|---|---|---|---|---|---|
| $H_{\text{eff}}^{(1,1)}$ |  |  | $560 \pm 10$ | $300 \pm 10$ | 1 | 1 | 4 by 4 |
| $H_{\text{eff}}^{(4,1)}$ | 3 | 0.89 | $560 \pm 10$ | $380 \pm 10$ | 4 | 1 | 16 by 16 |
| $H_{\text{eff}}^{(4,3)}$ |  |  | $560 \pm 10$ | $400 \pm 10$ | 4 | 3 | 32 by 32 |

TABLE II. Basis encoding used in $(N_{\max}, L_{\max}) = (1,1)$. Many-qubit states are written as $|q_3 q_2 q_1 q_0\rangle$ for direct encoding and $|q_1 q_0\rangle$ for compact encoding.

|  | $n$ | $m$ | $l$ | $s$ | $\bar{s}$ | Direct encoding | Compact encoding |
|---|---|---|---|---|---|---|---|
| ① | 0 | 0 | 0 | 1/2 | -1/2 | $|0001\rangle$ | $|00\rangle$ |
| ② | 0 | 0 | 0 | -1/2 | 1/2 | $|0010\rangle$ | $|01\rangle$ |
| ③ | 0 | 0 | 1 | 1/2 | -1/2 | $|0100\rangle$ | $|10\rangle$ |
| ④ | 0 | 0 | 1 | -1/2 | 1/2 | $|1000\rangle$ | $|11\rangle$ |

Besides the optimizers shown in the figure, we have also looked at other optimizers and presented the complete summary in Table. IV. For ideal simulation with the SV simulator, all results are in good agreement with the exact solution. In particular, the LBFGSB and COBYLA optimizers are used for the SV simulator in both direct encoding and compact encoding, and they quickly converge to the expected ground state energy. Typically, the LBFGSB optimizer works best for the SV simulator, reaching converged mass values at a much faster rate. For the QASM simulators that mimics the ideal quantum computer, we use both the COBYLA and SPSA optimizers, where the SPSA optimizer provides slightly better results than COBYLA. Due to sampling error from a measurement of 8,192 shots, the QASM results are much noisier and take longer to converge. The LBFGSB optimizer is also considered but fails to reach the expected ground state energy, as the LBFGSB optimizer depends on derivatives of the expectation values and does not perform sufficiently well with the inclusion of the sampling noise. Lastly, for the quantum computation performed on IBMQ Manila, the QNSPSA optimizer outperformed all the other optimizers (COBYLA, LBFGSB, SLSQP, SPSA), which is expected as it is tailored to the additional quantum noises [64] presented on a quantum device.

#### 2. Results of SSVQE

By using compact encoding and HEA, we present simulation results of our SSVQE approach to obtain the spectroscopy using compactly-encoded Hamiltonian operators $H_{\text{compact}}^{(1,1)}$, $H_{\text{compact}}^{(4,1)}$ and $H_{\text{compact}}^{(4,3)}$. In the case of $N_{\max} = L_{\max} = 1$, the 4-by-4 Hamiltonian matrix is mapped onto two qubits. We prepare the four orthogonal reference states $|00\rangle, |01\rangle, |10\rangle, |11\rangle$, and then evolve them via `EfficientSU2` ansatz with two repetition layers (12 parameters in total). For the cost function, we choose the weight vector $\vec{\omega} = (1.0, 0.5, 0.25, 0.125)$ such that

$$\mathcal{C}_{\vec{\omega}} = 1.0\, E_{|00\rangle} + 0.5\, E_{|01\rangle} \\ + 0.25\, E_{|10\rangle} + 0.125\, E_{|11\rangle}, \quad (24)$$

where $E_{|s\rangle} = \langle s|\hat{U}(\vec{\theta})|H_{\text{compact}}|\hat{U}(\vec{\theta})|s\rangle$ is the evolved expectation value for each orthogonal state $|s\rangle$. It is important to note that the respective reference state will be evolved in the order of its specified weight factor, namely, $|00\rangle$ becomes the ground state $(E_0)$, $|01\rangle$ becomes the first

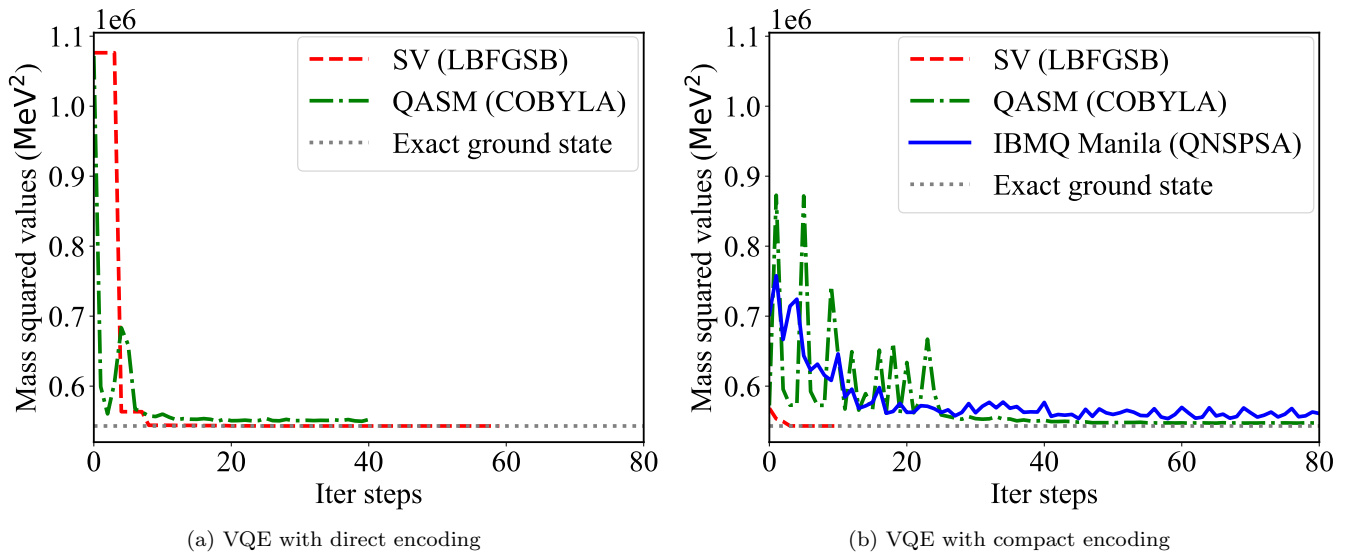

(a) VQE with direct encoding

(b) VQE with compact encoding

FIG. 3. The ground state energy at $N_{\max} = L_{\max} = 1$ calculated with the VQE approach using (a) direct encoding and (b) compact encoding. In each subfigure, we show results of the SV, QASM simulators, and IBMQ Manila quantum computing backend. The exact mass squared ground state of 543058.61 MeV$^2$ is also provided for comparison. The parenthesis behind each backend indicates its best respective optimizer used in the VQE optimization. Termination of each curve indicates the convergence of the expectation value by its respective optimizer.

TABLE III. Summary of quantum circuits used in direct encoding (UCC ansatz) and compact encoding (HEA) after transpilation to the IBM Manila basis-gate set { ID, X, RZ, SX, CX } with the highest optimization available in Qiskit 0.19.2.

| Ansatz | Qubits | Circuit depth | # Parameters | # Single-qubit gates | # CX-gates |
|--------|--------|---------------|--------------|----------------------|------------|
| UCC    | 4      | 67            | 3            | 73                   | 20         |
| HEA    | 2      | 11            | 8            | 20                   | 1          |

TABLE IV. Summary of VQE results using various backends (simulators and quantum device). Measurements from a total of 8,192 shots are included except for the SV results, along with their statistical uncertainties from measurements. The ground state energies (truncated to the nearest integers) in the table are in units of MeV$^2$. Note that algorithmic iterations can have different meanings to different optimizers, and the direct-encoding IBMQ Manila simulation did not converge in our simulation.

| Backend | Encoding | Optimizer | Ground state energy (MeV$^2$) | Iterations |
|---------|----------|-----------|-------------------------------|------------|
| SV | Direct | LBFGSB | 543059 | 60 |
|    | Direct | COBYLA | 543059 | 90 |
|    | Compact | LBFGSB | 543059 | 11 |
|    | Compact | COBYLA | 543059 | 344 |
| QASM | Direct | COBYLA | $552344 \pm 996$ | 41 |
|      | Direct | SPSA | $545767 \pm 152$ | 1051 |
|      | Compact | COBYLA | $547405 \pm 211$ | 99 |
|      | Compact | SPSA | $543065 \pm 6$ | 1551 |
| IBMQ Manila | Direct | QNSPSA | $1181783 \pm 11381$ | 200 |
|             | Compact | QNSPSA | $554568 \pm 1179$ | 200 |
| Exact solution | - | - | 543059 | - |

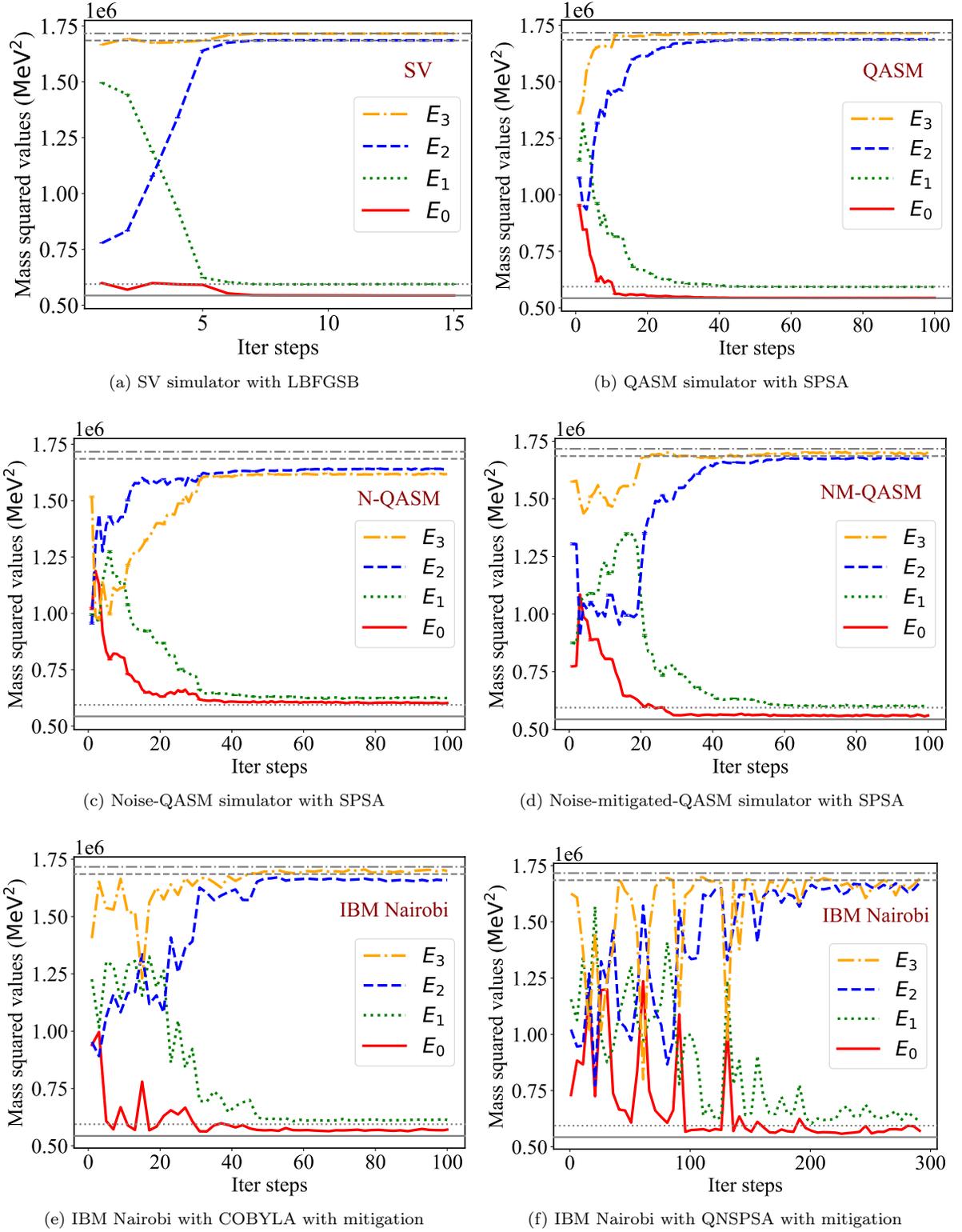

FIG. 4. The full mass spectroscopy at $N_{\max} = L_{\max} = 1$ calculated with the SSVQE approach using (a) SV simulator , (b) QASM simulator, (c) noise-QASM simulator, (d) noise-mitigated-QASM simulator, and (e,f) IBM Nairobi quantum computing backend. The four curves in each plot represent the evolution of the expectation values for the four orthogonal reference states throughout the optimization. The grey solid, dotted, dashed, and dot-dashed horizontal lines represent their respective exact energies $E_0$, $E_1$, $E_2$ and $E_3$ from solving the Hamiltonian directly on a classical computer. Best-performing optimizers are used for each optimization, and all measurements used 20,000 shots besides SV.

excited state ($E_1$), and so forth:

$$E_{|00\rangle} \to E_0, \; E_{|01\rangle} \to E_1,$$
$$E_{|10\rangle} \to E_2, \; E_{|11\rangle} \to E_3. \quad (25)$$

In Fig. 4, we present classically-simulated SSVQE results using local simulators (SV, QASM, QASM with noise model), as well as results from the quantum backend, IBM Nairobi, a recently-released 7-qubit quantum computer. All simulations used the same randomly-picked initial parameters for the ansatz and have the maximum number of shots allowed by the backend Nairobi (20,000). The upper two panels (a) and (b) show that spectroscopy obtained from the SSVQE simulation agrees very well with the exact energies using ideal simulators with and without statistical noise. In panel (c), we adopt a noise model based on IBM Nairobi using `NoiseModel` module from Qiskit, and we are able to estimate the effect of realistic quantum noises present for practical NISQ computers on the SSVQE approach. In panel (d), by using calibration techniques to mitigate the readout error, we demonstrate these quantum errors in panel (c) can be effectively controlled. Noise models are powerful tools that allow us to project quantum simulations onto realistic backends. In the last two panels (e-f), we show the quantum simulation (with error mitigation) on IBM Nairobi backends using the COBYLA and QN-SPSA optimizer respectively. Despite the variations in convergence pattern for each optimizer, both simulation results reach reasonable agreement with the true spectrum, and are also aligned with noise-mitigated-QASM results in panel (d). Note that re-calibration at a fixed interval (90 minutes) was necessary for the simulation as SSVQE optimization sometimes took 3-4 days to finish on IBM quantum backends (at the time this work was performed). In all, with a single optimization protocol, we find the SSVQE approach is capable of obtaining the complete spectroscopy for the hadron on simulators and quantum computers. The detailed information of these states for each simulation are summarized in Table. V.

Furthermore, we extend the SSVQE application to the larger Hamiltonian $H_\text{eff}^{(4,1)}$ and $H_\text{eff}^{(4,3)}$. With compact encoding, we map the 16-by-16 and 32-by-32 Hamiltonians onto four qubits and five qubits respectively. For $H_\text{eff}^{(4,1)}$, four orthogonal reference states, $|0000\rangle, |0001\rangle, |0010\rangle, |0011\rangle$, are prepared and evolved using the six-layer HEA (56 parameters in total). For $H_\text{eff}^{(4,3)}$, two orthogonal states, $|00000\rangle, |00001\rangle$ are prepared and evolved using five-layer HEA (60 parameters in total). We choose a similar cost function as shown previously in Eq. (24) and expect the reference states to evolve their corresponding energies in the spectrum specified by their respective weight coefficients. We apply the SSVQE approach using quantum simulators given limited available resources in carrying out the optimization iterations on the currently-available IBM quantum computers.

For the Hamiltonian $H_\text{eff}^{(4,1)}$, the SSVQE simulation results are presented in Fig. 5 with both the QASM and noise-mitigated-QASM simulator. Despite having a much more complicated ansatz we are able to obtain reasonable results compared to the exact spectroscopy given sufficient iterations. Note that we only present the lowest two states in the noise-mitigated-QASM simulation for the limited quantum backend (IBM Nairobi) mimicked by our noise model. The results for $H_\text{eff}^{(4,3)}$ are presented in Fig. 6 with both the QASM and noise-mitigated-QASM simulator. We can see the increased number of iterations and oscillatory pattern needed for convergence. In general, we find the results from ideal SV and QASM simulator agree with the exact energies, while the results from noise simulators are consistently greater than the exact energies due to quantum noises. For the same reason as well as long iterations for convergence, we did not run the SSVQE optimization for the two larger Hamiltonians on IBM quantum computers. Detailed numerical results for each state from both sets of simulations ($H_\text{eff}^{(4,1)}$ and $H_\text{eff}^{(4,3)}$) are presented in Table. V, where the SV simulation results are also included.

Lastly, we have tested the SSVQE and VQE simulations at the highest $N_\text{max} = 8$ cutoff as in the original problem, where a total of 7 qubits is needed with compact encoding for the 128-by-128 Hamiltonian. Although we were able to produce results in agreement with classical calculations using the SV simulator, we found that the simulation time per each cost evaluation in the QASM simulator (with and without noise models) increases nearly exponentially with the number of qubits. Together with the demand for an increased number of total iterations, an excessive amount of time would be required to achieve convergence with the QASM simulator.

### C. Light-front wave function as encoded quantum state

Light-front wave functions (LFWFs), cornerstones of the light-front Hamiltonian approach, enable us to calculate various physical quantities of interest and study the evolution of the system. As a result of the SSVQE optimization, we obtain the set of all the wave functions encoded on the quantum state directly. Using the `DensityMatrix` module we can obtain the density matrix of the quantum state representing its associated bound state in the spectrum.

In Fig. 7, we show the density matrix, $D_{ij} = |\psi_i\rangle \langle\psi_j|$, using the Hinton diagrams from Qiskit, of the lowest two states, the pion and rho meson obtained from optimizing the $H_\text{eff}^{(1,1)}$ Hamiltonian. In the left column, panels (a,c,e), we show the density matrices of the pion from the SV, QASM simulators, and IBM Nairobi. Note that the density matrix obtained in the SV simulator has exceedingly small imaginary parts to the naked eye, since they are ideal shot-free simulations that are closest to the classi-

TABLE V. Summary of SSVQE spectroscopy results using the SV, QASM, noise-mitigated-QASM (NM-QASM) simulators, and IBM Nairobi quantum computer. For IBM Nairobi, both simulation results using the COBYLA (left) and QNSPSA (right) optimizer are provided. The exact energies of the spectroscopy are provided as a reference. Measurements from a total of 20,000 shots are applied except for the SV results. All mass energies (truncated to the nearest integer) in the table are in units of $\text{MeV}^2$, along with their statistical uncertainties from measurements whenever available.

| $N_{\max}$ | $L_{\max}$ | Init state | Exact | SV | QASM | NM-QASM | IBM Nairobi | |
|---|---|---|---|---|---|---|---|---|
| 1 | 1 | $\lvert 00\rangle$ | 543059 | 543059 | $543661 \pm 40$ | $555448 \pm 795$ | 570482 | 571106 |
|   |   | $\lvert 01\rangle$ | 593915 | 593915 | $593427 \pm 39$ | $602433 \pm 832$ | 612433 | 613577 |
|   |   | $\lvert 10\rangle$ | 1685209 | 1685209 | $1687068 \pm 53$ | $1671575 \pm 854$ | 1659565 | 1674709 |
|   |   | $\lvert 11\rangle$ | 1716743 | 1716743 | $1714871 \pm 54$ | $1705904 \pm 749$ | 1698240 | 1692378 |
| 4 | 1 | $\lvert 0000\rangle$ | 369016 | 369256 | $373554 \pm 4133$ | $485813 \pm 4420$ | | |
|   |   | $\lvert 0001\rangle$ | 575707 | 576234 | $586963 \pm 3981$ | $642444 \pm 4267$ | | |
|   |   | $\lvert 0010\rangle$ | 737759 | 739282 | $786290 \pm 4195$ | | | |
|   |   | $\lvert 0011\rangle$ | 976608 | 981089 | $979853 \pm 4040$ | | | |
| 4 | 3 | $\lvert 00000\rangle$ | 336927 | 344136 | $337874 \pm 5683$ | $721627 \pm 7237$ | | |
|   |   | $\lvert 00001\rangle$ | 581652 | 595971 | $600335 \pm 5357$ | $957360 \pm 6935$ | | |

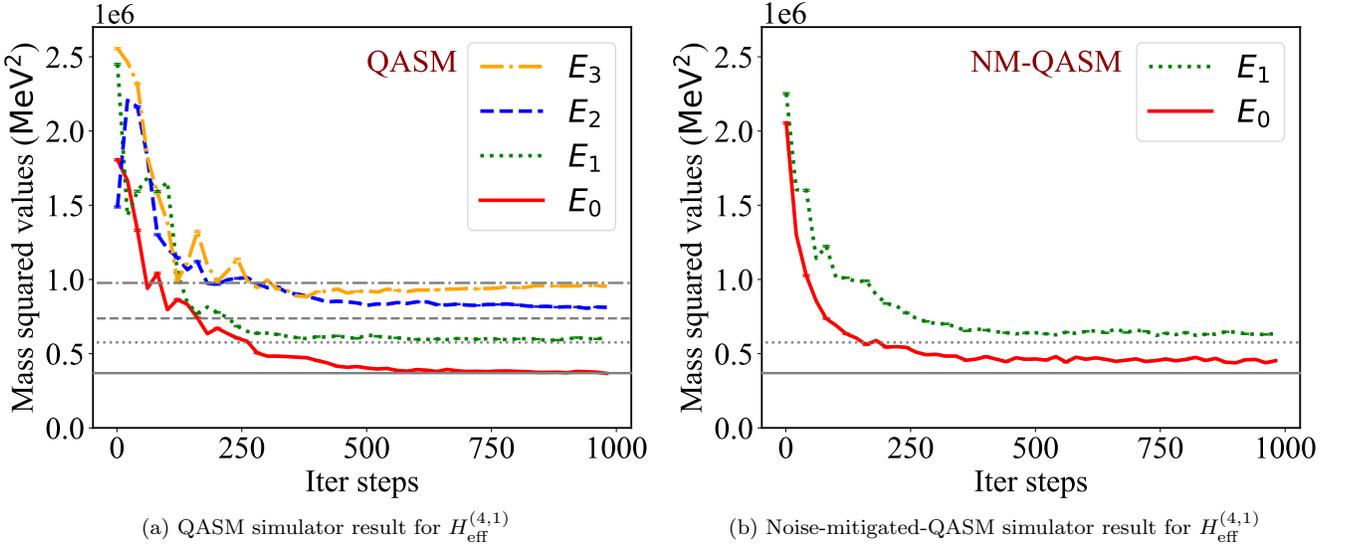

(a) QASM simulator result for $H_{\text{eff}}^{(4,1)}$

(b) Noise-mitigated-QASM simulator result for $H_{\text{eff}}^{(4,1)}$

FIG. 5. The low-lying mass spectroscopy calculated with the SSVQE approach using the QASM simulator (left panel) and noise-mitigated-QASM simulator (right panel) at $N_{\max} = 4, L_{\max} = 1$. The curves in each plot, four in panel (a) and two in panel (b), represent the evolution of the expectation values for the orthogonal reference states throughout the optimization. The grey solid, dotted, dashed, and dot-dashed horizontal lines represent the respective exact energies $E_0$, $E_1$, $E_2$ and $E_3$ from solving the Hamiltonian directly on a classical computer. The SPSA optimizers are used for both simulations with a measurement of 20,000 shots.

cal Hamiltonian diagonalization approaches. Therefore, we use the SV result as a reference density matrix for the simulation. For the QASM and IBM Nairobi density matrices, we can see the effects of statistical uncertainties and the effects of quantum noise respectively, which are aligned with our expectations. Similar trends can also be observed in the right column, panels (b,d,f), for the rho meson, despite of the difference in basis contributions.

In all these simulations, the trace of the density matrix and of the square of the density matrix are checked and always equal to unity up to numerical tolerance. In addition and most importantly, the orthogonality of the reference states are preserved throughout the simulation, from the exact simulation via SV simulator to quantum simulation via the IBM Nairobi noise model, which confirms how unitary evolution conserves the orthogonality of the states, a key feature of the SSVQE approach. The same observations are found for higher-dimensional Hamiltonians as well. Density matrices plotted using the Hinton diagrams are useful visualization tools that intuitively demonstrate the basis contributions of each hadron state as well as measurement/quantum noise in the simulations.

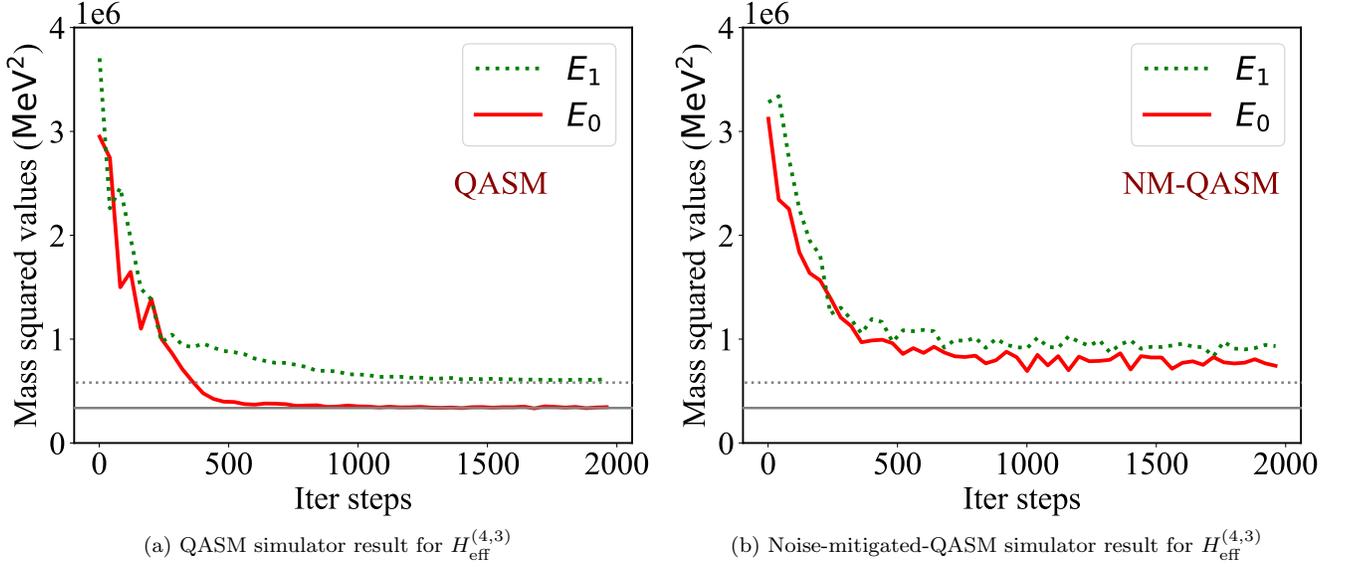

FIG. 6. The low-lying mass spectroscopy calculated with the SSVQE approach using the QASM simulator (left panel) and noise-mitigated-QASM simulator (right panel) at $N_{\max} = 4, L_{\max} = 3$. The two curves in each plot represent the evolution of the expectation values for the two orthogonal reference states throughout the optimization. The grey solid and dotted horizontal lines represent the exact respective energies $E_0$ (ground state) and $E_1$ (first excited state) from solving the Hamiltonian directly on a classical computer. The SPSA optimizers are used for both simulations with a measurement of 20,000 shots.

### D. Decay constants

Decay constants are experimentally important quantities and are defined as the local vacuum-to-hadron matrix element of the quark current operators. By taking the "+" current component and the $m_j = 0$ state of the meson [21], the pseudoscalar decay constants ($f_\mathrm{P}$) and vector meson decay constants ($f_\mathrm{V}$) in the BLFQ basis function are written as

$$f_{\mathrm{P,V}} = \sqrt{2N_c} \int_0^1 \frac{\mathrm{d}x}{\sqrt{x(1-x)}} \int \frac{\mathrm{d}^2 \boldsymbol{k}_\perp}{(2\pi)^3} \psi^{(m_j=0)}_{\uparrow\downarrow\mp\downarrow\uparrow}(x, \boldsymbol{k}_\perp)$$

$$\equiv \frac{\kappa \sqrt{N_c}}{\pi} \sum_{nl} (-1)^n C_l(m_q, \kappa)$$

$$\times \left( \tilde{\psi}^{(m_j=0)}_{\uparrow\downarrow}(n,0,l) \mp \tilde{\psi}^{(m_j=0)}_{\downarrow\uparrow}(n,0,l) \right), \quad (26)$$

where $N_c = 3$, $C_l$ is the resulting coefficient that depends on $m_q^2/\kappa$ [69], and $\tilde{\psi}^{(m_j=0)}_{s\bar{s}}$ is the basis coefficient of the LFWF defined in Eq. (5). In this case, the decay constants are linear with the LFWF, $f_{\mathrm{P,V}} \propto \langle \nu_{\mathrm{P,V}} | \psi(\vec{\theta}) \rangle$, for some vector $\nu_{\mathrm{P,V}}$ which depends on the specific LFWF basis encoding on the qubits. To measure the decay constant directly on quantum computers [9], we construct the Pauli operators from $|\nu\rangle\langle\nu|$, such that

$$|\langle \nu_{\mathrm{P,V}} | \psi(\vec{\theta}) \rangle| = \sqrt{\langle \psi(\vec{\theta}) | (|\nu_{P,V}\rangle\langle\nu_{P,V}|) | \psi(\vec{\theta}) \rangle}, \quad (27)$$

and then map $|\nu_{P,V}\rangle\langle\nu_{P,V}|$ onto qubits by compact encoding to obtain the decay constant operators $|\nu_{P,V}\rangle\langle\nu_{P,V}|_q$. Therefore, decay constants can be evaluated directly on the quantum computer as the expectation value of the $|\nu_{P,V}\rangle\langle\nu_{P,V}|_q$ operator on the specified final state.

For $N_{\max} = L_{\max} = 1$, according to Table II, $\nu_\mathrm{P}^{(1,1)} = (1, -1, 0, 0)$ and $\nu_\mathrm{V}^{(1,1)} = (1, 1, 0, 0)$, each corresponding to the singlet and triplet LFWFs in Eq. (26) respectively. By mapping the vectors to qubits, we have

$$|\nu_\mathrm{P}^{(1,1)}\rangle\langle\nu_\mathrm{P}^{(1,1)}|_q = 0.5 \, (II - IX + ZI - ZX), \quad (28)$$

$$|\nu_\mathrm{V}^{(1,1)}\rangle\langle\nu_\mathrm{V}^{(1,1)}|_q = 0.5 \, (II + IX + ZI + ZX), \quad (29)$$

where P stands for the pseudoscalar meson and V stands for the vector meson. The decay constant operators for $H_\mathrm{eff}^{(4,1)}$ are more involved and are included in Appendix C.

In the SSVQE spectroscopy, the lowest two states are identified as the pseudoscalar meson $\pi$ and the vector meson $\rho$. With their respective evolved final states, their decay constants, $f_\pi$ (130 MeV) and $f_\rho$ (216 MeV), are measured as the expectation value of $|\nu\rangle\langle\nu|$ using Eq. (27) and presented in Table VI for various simulators and quantum computers. By taking sampling error into account, we can see that the obtained decay constants from the SV, QASM, and NM-QASM simulators are in reasonable agreement with the exact calculation. It is important to point out that despite the considerable difference in spectroscopy, the NM-QASM results for decay constants at $N_{\max} = 4, L_{\max} = 3$ are in the close agreement with exact data. With our limited basis size, the SSVQE approach proves to be a useful tool in analyzing hadronic structures such as decay constants of low-lying states given successful optimization of the spectroscopy.

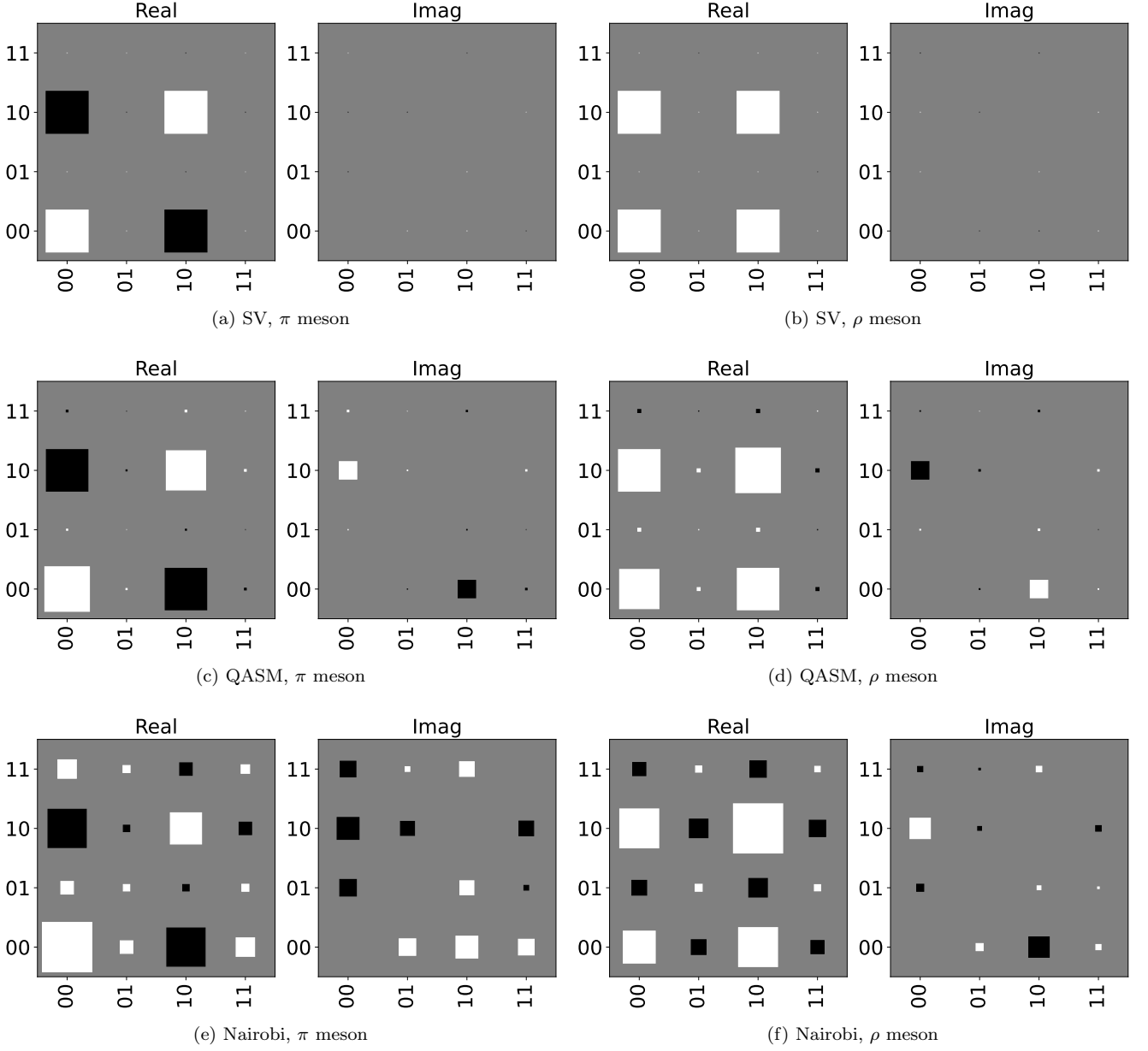

FIG. 7. Visualization of selected density matrices (using the Hinton diagrams) of the pion and the rho meson obtained from the SV, QASM simulators, and IBM Nairobi for the $H_{\text{eff}}^{(1,1)}$ Hamiltonian. The density matrix of each state, a matrix of complex numbers, is represented by a real ("Real") and a imaginary ("Imag") diagram. Here, the white/black boxes represent the positive/negative amplitudes of their corresponding basis. Their sizes, or box areas, represent the strength of the amplitudes.

### E. Parton distribution function

The parton distribution function (PDF) is another important experimentally-accessible quantity that is often discussed in the context of QCD scale evolution. It describes the probability of finding a particle with longitudinal momentum fraction $x$ at some factorization scale $\mu$ related to the experimental conditions. In the BLFQ basis representation [20, 70], the PDF for finding a quark

TABLE VI. Summary of decay constants for $\pi$ and $\rho$ by measuring final states obtained from the SSVQE results using the SV, QASM, noise-mitigated-QASM (NM-QASM) simulators, and IBM Nairobi quantum computers. Decay constants in the table are in units of MeV, and their statistical errors are provided from a measurement of 20,000 shots except for the SV simulator. The experimental decay constants for $\pi$ and $\rho$ are 130 MeV and 216 MeV respectively according to the PDG data [67]. The decay constant result of IBM Nairobi uses the optimized parameters from the COBYLA optimizer.

|  | $N_{\max}$ | $L_{\max}$ | Exact result | SV | QASM | NM-QASM | IBM Nairobi |
|---|---|---|---|---|---|---|---|
| $f_\pi$ | 1 | 1 | 178.18 | 178.18 | $177.11 \pm 4.94$ | $174.64 \pm 6.61$ | $164.20 \pm 8.51$ |
| $f_\rho$ |   |   | 178.18 | 178.18 | $177.17 \pm 4.88$ | $174.55 \pm 6.65$ | $167.76 \pm 8.21$ |
| $f_\pi$ | 4 | 1 | 199.36 | 200.61 | $200.32 \pm 11.99$ | $196.02 \pm 12.23$ |  |
| $f_\rho$ |   |   | 227.63 | 230.08 | $228.13 \pm 10.10$ | $224.80 \pm 10.55$ |  |
| $f_\pi$ | 4 | 3 | 199.34 | 199.57 | $201.90 \pm 10.72$ | $186.15 \pm 11.01$ |  |
| $f_\rho$ |   |   | 229.25 | 230.04 | $228.58 \pm 9.58$ | $203.04 \pm 10.58$ |  |

in the meson system, is expressed as

$$q(x;\mu) = \frac{1}{x(1-x)} \sum_{s\bar{s}} \int \frac{\mathrm{d}^2 \boldsymbol{k}_\perp}{2(2\pi)^3} |\psi_{s\bar{s}}^{(m_j=0)}(x, \boldsymbol{k}_\perp)|^2$$

$$\equiv \frac{1}{4\pi} \sum_{s\bar{s}} \sum_{nm} \sum_{l\bar{l}} \tilde{\psi}_{s\bar{s}}^{*(m_j=0)}(n, m, \bar{l})$$

$$\times \tilde{\psi}_{s\bar{s}}^{(m_j=0)}(n, m, l) \chi_l(x) \chi_{\bar{l}}(x), \quad (30)$$

where $\tilde{\psi}_{s\bar{s}}^{(m_j=0)}$ is the basis coefficient of the LFWF defined in Eq. (5) and the PDF satisfies the normalization $\int_0^1 q(x)\, \mathrm{d}x = 1$. The truncation to a Fock space with a single quark and anti-quark implies the model Hamiltonian is appropriate to a factorization scale typically much lower than the scale accessed in high-energy experiments that measure the PDF.

Various approaches [17, 71] can be adopted in the calculation of the PDF on a quantum computer. In this work, we take advantage of the BLFQ basis formulation to decompose the finite sum expression in Eq. (30) and evaluate each term respectively by using projection operator $\hat{U}_\mathrm{p}(s, \bar{s}, n, m, l)$ that map the quantum state into the corresponding basis,

$$q(x) = \sum_{s\bar{s}} \sum_{nm} \sum_{l\bar{l}} \langle \psi(\vec{\theta})| \hat{O}_\mathrm{pdf}(x) |\psi(\vec{\theta})\rangle, \quad (31)$$

$$\hat{O}_\mathrm{pdf}(x) = \hat{U}_\mathrm{p}(s, \bar{s}, n, m, \bar{l})^\dagger \hat{U}_\mathrm{p}(s, \bar{s}, n, m, l)$$

$$\times \frac{\chi_l(x) \chi_{\bar{l}}(x)}{4\pi}. \quad (32)$$

Here, $\hat{O}_\mathrm{pdf}(x)$ is the unitary operator to evaluate each subterm contribution of the PDF at a given longitudinal momentum fraction $x$. The PDF operator is then mapped onto the qubits. In Appendix D, we present examples of the qubitized PDF operators at $x = 0.5$ and $x = 0.25$.

By taking the lowest two states, the $\pi$ and $\rho$ mesons, obtained from SSVQE optimization, we show the calculation of PDFs from QASM simulators in Fig. 8, following Eq. (31), for all three Hamiltonians considered in this work. These obtained PDFs are sampled at 19 evenly-spaced longitudinal momentum fractions and they are in reasonable agreement with those from the exact classical results. Going through panels (a-c), one can see the sensitivity of the PDF to the model parameters, since different quark mass $m_f$ is used. From panels (a) and (b), it is important to see that the PDFs for the pseudoscalar and the vector mesons are almost identical due to the lack of longitudinal excitation modes in both truncations, i.e., $L = 1$. Comparing panels (b) and (c), we can see the $\pi$ and $\rho$ mesons are sensitive to different longitudinal excitation modes as expected from solving $H_{\max}^{(4,1)}$ and $H_{\max}^{(4,3)}$ directly. The results from SV simulators are generally omitted because they are almost identical to the exact calculations; however, we provide panel (d) showing the PDF obtained from the SV simulator at $N_{\max} = 4, L_{\max} = 3$ to demonstrate that shot-free ideal simulation is able to obtain perfect agreement with the exact PDFs at the largest basis while QASM simulation starts to have difficulty due to statistical uncertainty. Lastly, the PDFs from noise-mitigated-QASM simulations are similar to the QASM simulations shown in panels (a-c) except for larger uncertainty bars.

## V. SUMMARY AND DISCUSSIONS

In this work, we used the variational quantum eigensolver (VQE) and the subspace-search variational quantum eigensolver (SSVQE) to study the hadron structures of the light meson system within the basis light-front quantization (BLFQ) approach. Our model Hamiltonian was taken from a previous work with fitted parameters obtained for three reduced basis spaces, $(N_{\max}, L_{\max}) = (1, 1)$, $(N_{\max}, L_{\max}) = (4, 1)$, and $(N_{\max}, L_{\max}) = (4, 3)$. Mass spectroscopy, decay constants, and parton distribution functions were directly calculated by using the VQE/SSVQE approach on the quantum circuits, using

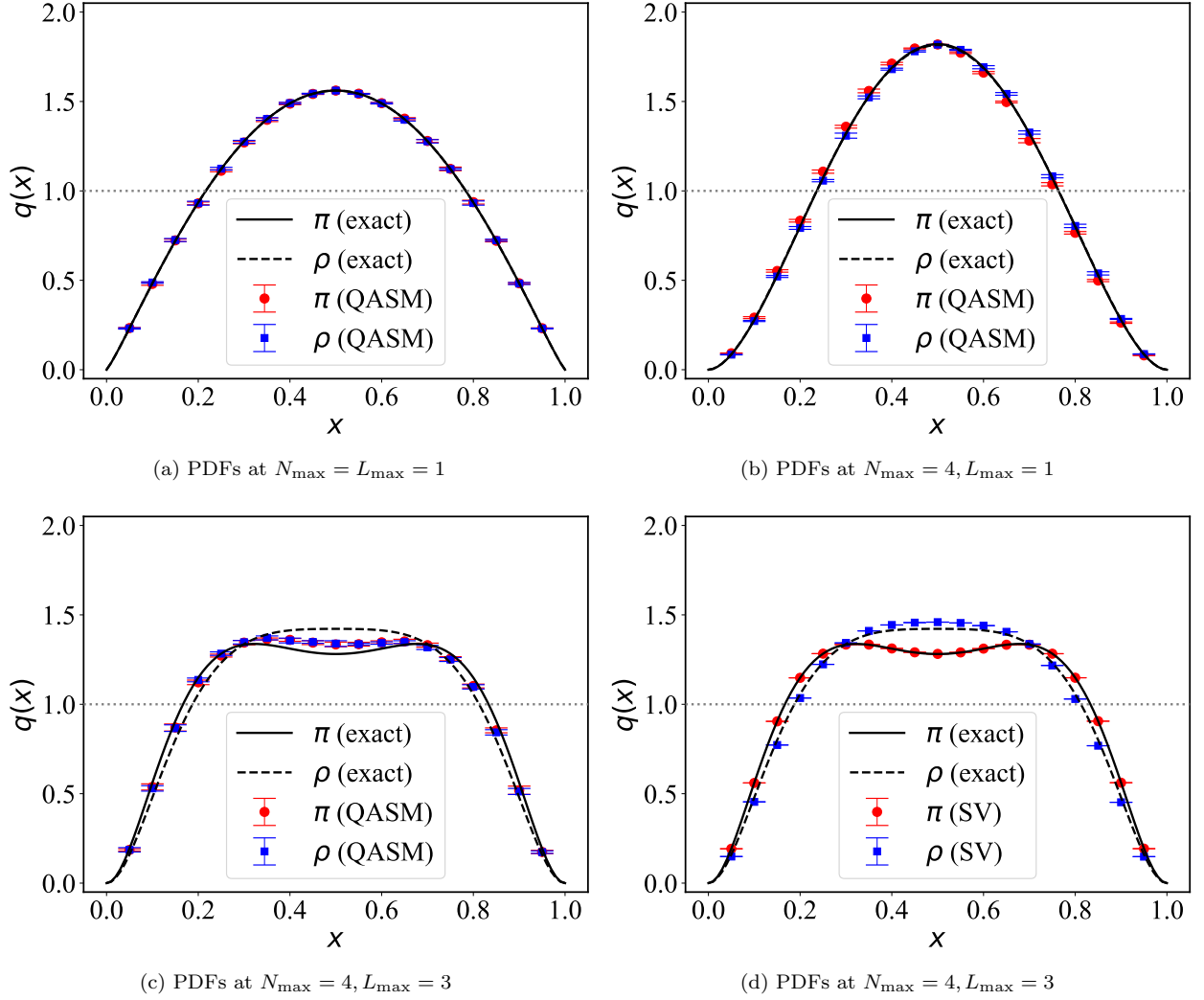

FIG. 8. PDFs calculated with the lowest two states, $\pi$ and $\rho$, obtained with the SSVQE approach at $(N_{\max}, L_{\max}) = (1,1)$, $(N_{\max}, L_{\max}) = (4,1)$, and $(N_{\max}, L_{\max}) = (4,3)$ using the QASM simulator. The solid(dashed) black curves represent the exact PDFs for the $\pi(\rho)$ mesons calculated on classical computers. The QASM simulated results of the PDFs are calculated at 19 evenly-spaced longitudinal momentum fractions and the sampling errors from a measurement of 20,000 shots are provided as their error bars respectively. The PDFs from ideal SV simulators on the largest basis are provided as a reference in panel (d).

various quantum simulators and IBM quantum computers.

For the VQE approach, we focused on the smallest nontrivial Hamiltonian with $(N_{\max}, L_{\max}) = (1,1)$, and used the UCC ansatz with direct encoding and the hardware efficient ansatz (HEA) with compact encoding to obtain the lowest energy state. Compact encoding with HEAs prove to be particularly useful when carried onto the currently-available NISQ quantum computers. For the SSVQE approach, we considered three Hamiltonians of increasing basis sizes: $(N_{\max}, L_{\max}) = (1,1)$, $(N_{\max}, L_{\max}) = (4,1)$, and $(N_{\max}, L_{\max}) = (4,3)$. In particular, we used the HEAs with compact encoding to obtain the lowest energy states in the spectroscopy. The mass eigenvalue results are generally consistent with the exact classical solution. In addition, by taking advantage of BLFQ basis functions and its qubit encoding, we computed the decay constants and the parton distribution function directly on the quantum circuits. Especially, we focused on the lowest two states, which correspond to the $\pi$ and $\rho$ mesons. For all of our simulated results, both the statevector (SV) and QASM simulators give reasonable agreement with the exact results. The noise-mitigated QASM simulators are useful for projecting simulations to quantum computers, and match exact results when basis sizes are relatively small. In terms of optimizers, we find COBYLA and LBFGSB optimizers perform best for SV simulators; while COBYLA and SPSA for QASM simulators, which is expected from the derivative-free optimizers. For superconducting NISQ devices, such as IBMQ Manila or IBM Nairobi, we found the QNSPSA and COBYLA perform the best among all the optimizers.

This work represents a first step to study hadron spectroscopy as well as observables within the BLFQ formalism on quantum computers. The VQE/SSVQE approaches prove to be particularly useful tools for basis Hamiltonian formalisms. Unlike classical computation, we are using the quantum state itself to encode classical information in a quantum simulation. With the exponential state space provided by the quantum system, the quantum state itself is efficient as it only needs a logarithmic amount of resources, as seen in the compact encoding of the Hamiltonian. In the future, we expect to include higher basis state contributions for a more accurate description of the light meson bound-state problem, provided that better noise mitigation or correction methods are implemented. We also plan to carry out the calculations on more robust quantum devices with higher quantum volume when they become accessible. When larger computations become possible, we will further investigate practical strategies to mitigate variational optimization problems such as barren plateaus. Lastly, we anticipate extending our work to compute other important hadronic properties such as transition amplitudes, which are feasible within the SSVQE approach.

## ACKNOWLEDGEMENTS

We wish to thank M. Kreshchuk, M. Li, M. Eshaghian-Wilner, C. Salgado, and A. Khokhar for valuable discussions. We acknowledge the use of IBM Quantum services for this work. The views expressed are those of the authors, and do not reflect the official policy or position of IBM or the IBM Quantum team. This work is supported by the US Department of Energy (DOE), Office of Science, under Grant No. DE-FG02-87ER40371. Computational resources were provided by the National Energy Research Scientific Computing Center (NERSC), which is supported by the U.S. DOE Office of Science under Contract No. DE-AC02-05CH11231. WQ is also supported by Xunta de Galicia (Centro singular de investigacion de Galicia accreditation 2019-2022), European Union ERDF, the "Maria de Maeztu" Units of Excellence program under project CEX2020-001035-M, the Spanish Research State Agency under project PID2020-119632GB-I00, and European Research Council under project ERC-2018-ADG-835105 YoctoLHC.

## Appendix A: Light-front coordinates

The light-front coordinates are defined as $x^\mu = (x^+, x^-, x^1, x^2)$, where $x^+ = x^0 + x^3$ is the light-front time, $x^- = x^0 - x^3$ is the longitudinal coordinate, $\boldsymbol{x}^\perp = (x^1, x^2)$ are the transverse coordinates. The corresponding metric tensor and its inverse are

$$g_{\mu\nu} = \begin{pmatrix} 0 & 1/2 & 0 & 0 \\ 1/2 & 0 & 0 & 0 \\ 0 & 0 & -1 & 0 \\ 0 & 0 & 0 & -1 \end{pmatrix}, \tag{A1}$$

$$g^{\mu\nu} = \begin{pmatrix} 0 & 2 & 0 & 0 \\ 2 & 0 & 0 & 0 \\ 0 & 0 & -1 & 0 \\ 0 & 0 & 0 & -1 \end{pmatrix}. \tag{A2}$$

## Appendix B: Pauli matrices

The Pauli matrices acting on the $i$-th qubit are defined as

$$X_i = \begin{pmatrix} 0 & 1 \\ 1 & 0 \end{pmatrix}, Y_i = \begin{pmatrix} 0 & -i \\ i & 0 \end{pmatrix}, Z_i = \begin{pmatrix} 1 & 0 \\ 0 & -1 \end{pmatrix}, \tag{B1}$$

where subscripts are sometimes omitted for simplicity. $I$ is used for the identity matrix.

## Appendix C: Decay constant operators

In the case of $N_{\max} = 4, L_{\max} = 1$, according to Table. VII, the vectors $\nu$ for the decay constants are defined as

$$\nu_{\mathrm{P}} = (1, -1, 0, 0, 0, 0, 0, 0, -1, 1), \tag{C1}$$
$$\nu_{\mathrm{V}} = (-1, -1, 0, 0, 0, 0, 0, 0, 1, 1), \tag{C2}$$

and the corresponding decay constant operators on the qubits in compact encoding are

$$\begin{aligned}&|\nu_{\mathrm{P}}\rangle\langle\nu_{\mathrm{P}}|_q = \\ &0.25 \Big( IIII - IIIX + IIZI - IIZX \\ &+ IZII - IZIX + IZZI - IZZX \\ &- XIII + XIIX - XIZI + XIZX \\ &- XZII + XZIX - XZZI + XZZX \Big), \end{aligned} \tag{C3}$$

$$\begin{aligned}&|\nu_{\mathrm{V}}\rangle\langle\nu_{\mathrm{V}}|_q = \\ &0.25 \Big( IIII + IIIX + IIZI + IIZX \\ &+ IZII + IZIX + IZZI + IZZX \\ &- XIII - XIIX - XIZI - XIZX \\ &- XZII - XZIX - XZZI - XZZX \Big). \end{aligned} \tag{C4}$$

## Appendix D: Parton distribution function operators

We present examples of the qubitized parton distribution function (PDF) operators $\hat{O}_{\mathrm{pdf}}(x)$ at $x = 0.5$ and

$x = 0.25$ (up to second decimal places) in compact encoding,

$$\hat{O}^{(1,1)}_{\text{pdf}}(0.5)_q = 1.30\,II - 1.29\,IX - 0.18\,IZ, \quad \text{(D1)}$$

$$\hat{O}^{(1,1)}_{\text{pdf}}(0.25)_q = 0.78\left(II + IZ\right), \quad \text{(D2)}$$

$$\hat{O}^{(4,1)}_{\text{pdf}}(0.5)_q = 0.39\left(IIII + IIIZ - ZZII - ZZIZ\right), \quad \text{(D3)}$$

$$\hat{O}^{(4,1)}_{\text{pdf}}(0.25)_q = 0.65\left(IIII - IIIX - ZZII + ZZIX\right) + 0.09\left(ZZIZ - IIIZ\right). \quad \text{(D4)}$$

## Appendix E: Hamiltonian operator for $N_{\max} = 4$, $L_{\max} = 1$

In the case of $N_{\max} = 4$, $L_{\max} = 1$, the Hamiltonian operator in the compact representation is

$$\begin{aligned}
H^{(4,1)}_{\text{compact}} &= 1980715\,IIII - 526128\,IIIZ + 495549\,IIXI \\
&+ 49226\,IIXZ - 545122\,IIZI + 11747\,IIZZ \\
&+ 30028\,IYIY - 22551\,IYXY + 28639\,IYYI \\
&+ 20978\,IYYX - 66\,IYYZ + 5575\,IYZY \\
&+ 2002\,XXII + 7851\,XXIZ - 374\,XXXI \\
&- 5044\,XXXZ + 698\,XXZI + 3286\,XXZZ \\
&+ 74640\,XZII - 56314\,XZIX + 7971\,XZIZ \\
&- 35556\,XZXI + 29701\,XZXX - 6380\,XZXZ \\
&+ 899\,XZYY + 18521\,XZZI - 14096\,XZZX \\
&+ 2675\,XZZZ + 30028\,YIIY - 22551\,YIXY \\
&- 28639\,YIYI + 20978\,YIYX + 66\,YIYZ \\
&+ 5575\,YIZY + 2002\,YYII + 7851\,YYIZ \\
&- 374\,YYXI - 5044\,YYXZ + 698\,YYZI \\
&+ 3286\,YYZZ - 74640\,ZXII - 56314\,ZXIX \\
&- 7971\,ZXIZ + 35556\,ZXXI + 29701\,ZXXX \\
&+ 6380\,ZXXZ + 899\,ZXYY - 18521\,ZXZI \\
&- 14096\,ZXZX - 2675\,ZXZZ + 237267\,ZZII \\
&- 29297\,ZZIZ + 58469\,ZZXI + 17304\,ZZXZ \\
&- 6135\,ZZZI - 8354\,ZZZZ \quad \text{(E1)}
\end{aligned}$$

and the corresponding basis identification is included in Table. VII.

TABLE VII. Basis encoding used in $(N_{\max}, L_{\max}) = (4, 1)$. Many-qubit states are written as $|q_3 q_2 q_1 q_0\rangle$.

|    | $n$ | $m$ | $l$ | $s$ | $\bar{s}$ | Compact encoding |
|----|-----|-----|-----|-----|-----------|------------------|
| 1  | 0 | 0  | 0 | 1/2  | -1/2 | $|0000\rangle$ |
| 2  | 0 | 0  | 0 | -1/2 | 1/2  | $|0001\rangle$ |
| 3  | 0 | 0  | 1 | 1/2  | -1/2 | $|0010\rangle$ |
| 4  | 0 | 0  | 1 | -1/2 | 1/2  | $|0011\rangle$ |
| 5  | 0 | 1  | 0 | -1/2 | -1/2 | $|0100\rangle$ |
| 6  | 0 | 1  | 1 | -1/2 | -1/2 | $|0101\rangle$ |
| 7  | 0 | -1 | 0 | 1/2  | 1/2  | $|0110\rangle$ |
| 8  | 0 | -1 | 1 | 1/2  | 1/2  | $|0111\rangle$ |
| 9  | 1 | 0  | 0 | 1/2  | -1/2 | $|1000\rangle$ |
| 10 | 1 | 0  | 0 | -1/2 | 1/2  | $|1001\rangle$ |
| 11 | 1 | 0  | 1 | 1/2  | -1/2 | $|1010\rangle$ |
| 12 | 1 | 0  | 1 | -1/2 | 1/2  | $|1011\rangle$ |
| 13 | 1 | 1  | 0 | -1/2 | -1/2 | $|1100\rangle$ |
| 14 | 1 | 1  | 1 | -1/2 | -1/2 | $|1101\rangle$ |
| 15 | 1 | -1 | 0 | 1/2  | 1/2  | $|1110\rangle$ |
| 16 | 1 | -1 | 1 | 1/2  | 1/2  | $|1111\rangle$ |